# Thermal Conductivity Measurements in Nanosheets via Bolometric Effect

Onur Çakıroğlu[2], Naveed Mehmood[1], Mert Miraç Çiçek[1], Aizimaiti Aikebaier[1], Hamid Reza Rasouli[1], Engin Durgun[1], T. Serkan Kasırga[1,2]

[1] Bilkent University UNAM – National Nanotechnology Research Center, Ankara, Turkey 06800

[2] Department of Physics, Bilkent University, Ankara, Turkey 06800

**Thermal conductivity measurement techniques for materials with nanoscale dimensions require fabrication of very complicated devices or their applicability is limited to a class of materials. Discovery of new methods with high thermal sensitivity are required for the widespread use of thermal conductivity measurements in characterizing materials' properties. We propose and demonstrate a simple non-destructive method with superior thermal sensitivity to measure the in-plane thermal conductivity of nanosheets and nanowires using the bolometric effect. The method utilizes laser beam heating to create a temperature gradient, as small as a fraction of a Kelvin, over the suspended section of the nanomaterial with electrical contacts. Local temperature rise due to the laser irradiation alters the electrical resistance of the device, which can be measured precisely. This resistance change is then used to extract the temperature profile along the nanomaterial using thermal conductivity as a fitting parameter. We measured the thermal conductivity of $V_2O_3$ nanosheets to validate the applicability of the method and found an excellent agreement with the literature. Further, we measured the thermal conductivity of metallic $2H-TaS_2$ for the first time and performed *ab initio* calculations to support our measurements. Finally, we discussed the applicability of the method on semiconducting nanosheets and performed measurements on $WS_2$ and $MoS_2$ thin flakes.**

Heat in solids is transferred via phonons and electrons. Contribution of each heat carrier to the overall thermal conductivity of a solid depends on several factors such as electrical conductivity of the material, impurities, defects, crystallinity and electronic correlations[1,2]. A precise measurement of the thermal conductivity in nano-sized materials is important as the heat removal has become a critical issue for the electronics industry and as the temperature dependent thermal properties can provide valuable insights to materials' characteristics such as the ones that result due to electronic correlations[3–5]. There are various steady-state and transient measurement techniques available for thermal conductivity measurements for nanosheets and nanowires. Micro-Raman thermometry[6–11] and microbridge method[12,13] are among the most commonly used steady-state methods for the thermal conductivity measurements. Time domain thermal reflectance (TDTR)[14–17], frequency domain thermal reflectance (FDTR)[18–20] and the 3ω method[21] are among the transient measurement methods. Each method has its own strengths and weaknesses over the others[22].

Raman thermometry is a commonly used method to measure the thermal conductivity of nanosheets. The technique relies on identification of the local temperature rise over the suspended part of the nanosheet by using the temperature dependent shift of a Raman peak. Then, solving the heat transport equation with the extracted average temperature rise gives a measure of the in-plane thermal conductivity. For the materials with Raman peaks that are not very sensitive to temperature variations or with broad featureless Raman spectra, applicability of the technique is limited[23]. For instance, low temperature thermal conductivity measurements for graphene cannot be performed via Raman thermometry as the shift in the 2D peak due to temperature change at low temperatures is beyond the measurement sensitivity[11,24]. Microbridge thermometry is a scaled down version of the absolute thermal conductivity measurement technique[22]. The material under investigation is suspended across the heating and sensing elements to extract the thermal conductivity. The major drawback is the tremendous difficulty with the fabrication of the nanostructures. Sample contamination is another issue with the multi-step processes performed to achieve the desired device structure particularly for atomically thin materials. Transient thermal conductivity measurement methods pose challenges in the analysis of the acquired data[25–27] and complexity of the measurement setups limits the applicability of the methods[22].

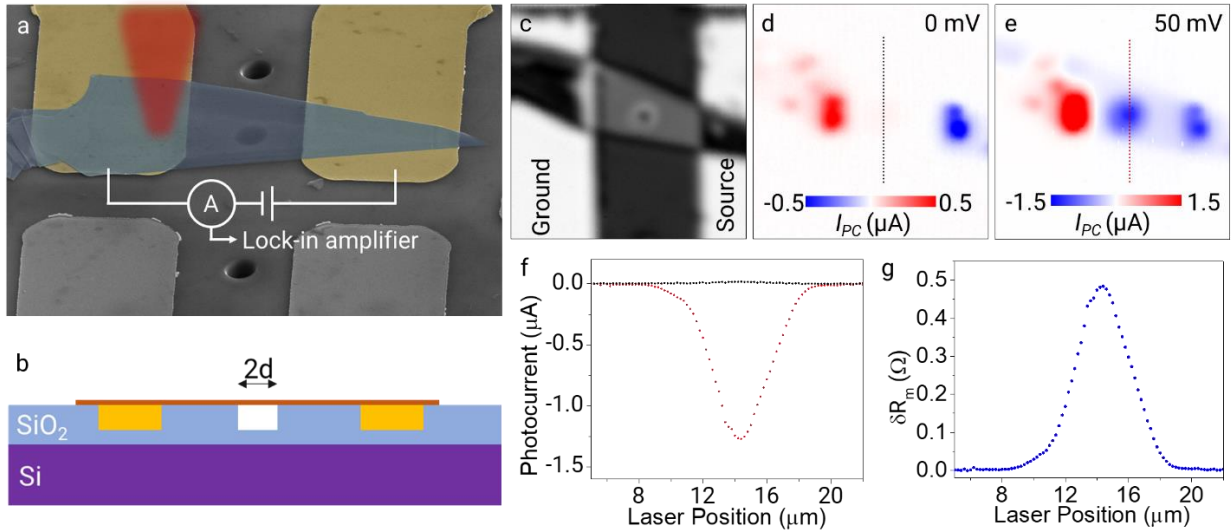

**Figure 1 a.** The proposed thermal conductivity measurement method is outlined on a false-colored SEM micrograph of a 2H-TaS$_2$ flake placed on the gold contacts. The red cone represents the focused scanning laser beam chopped at the frequency, $f$. A current pre-amplifier that acts as a virtual ground measures and amplifies the current due to applied dc bias and the current generated by the laser beam. Output of the measurement is fed to a lock-in amplifier referenced by the laser chopper. The measurement yields the change in the current due to the laser beam. **b.** Cross-sectional view of the device is depicted in the figure. Nanosheet, shown in orange, is overlaid on a hole of radius $d$ and on gold contacts, shown by yellow rectangles, patterned in SiO$_2$ with the top surfaces exposed. **c.** Scanning photocurrent microscopy (SPCM) reflection map of a device with source and ground contacts are labelled. **d.** Photocurrent map taken under 0 mV bias shows the Seebeck current generated at the contact-flake boundaries. **e.** Photocurrent map taken under 50 mV shows a negative photoresponse throughout the crystal and a distinct decrease around the hole. **f.** Line trace taken along the dashed lines in 0 and 50 mV scans show the photocurrent generated via the laser scan and **g.** shows the corresponding overall resistance change at each laser position.

In this paper, we introduce a novel method to measure the thermal conductivity of nanosheets based on the photothermally induced local electrical resistivity change, known as the bolometric effect. The electrical resistivity, $\rho(T)$, of materials have characteristic temperature dependency. In the linear approximation, metallic resistivity follows $\rho(T) = \rho_0 + \varrho(T - T_0)$. Here, $\rho_0$, $\varrho$ and $T_0$ are the resistivity at room temperature, temperature coefficient of resistivity and the room temperature, respectively. Similarly, a thermally activated resistivity, $\rho(T) = \rho_0 \exp(E_A/k_B T)$, can be defined for the semiconductors, where $E_A$ is the activation energy and $k_B$ is the Boltzmann constant. Any local source of heat will result in a thermal distribution over the suspended part of the nanosheet, $T(r; \kappa)$, depending on the thermal conductivity, $\kappa$, of the material. Using $T(r; \kappa)$, electrical resistivity for each point on the sample can be defined and the total resistance of the laser heated crystal can be calculated. Thus, a precise measurement of the photothermally induced electrical resistance change can be used to extract the thermal profile by using $\kappa$ as a fitting parameter. A similar method has been previously employed to measure the thermal conductivity of the single-walled carbon nanotube fibers in a much limited context[28].

To realize the theoretical scheme outlined above, we implemented the following experimental setup (Figure 1a-b). A commercial scanning photocurrent microscope (SPCM) is used for the measurements. SPCM is equipped with a 40x objective that focuses a laser beam to a Gaussian spot. The gold electrical contacts to the sample are patterned using a negative tone resist to prevent side wall formation after lift-off and deposited in to pits that are etched by the thickness of the gold to be deposited to avert the suspending of the thin flake (Figure 1b). Resistance measurements are performed to extract the electrical resistivity, $\rho(T)$ of the sample with dimensions measured via atomic force microscopy (AFM). Then, the contacts are used for the SPCM measurements. The laser beam chopped at a certain frequency ($f \approx 2$ kHz) scans the whole sample. Scanning the laser over the sample ensures that the laser will always pass through the center of the hole and the error due to the alignment of the laser spot with the hole will be minimized. When the laser beam passes over the hole of a radius $d$ etched under the nanosheet, laser heating induced resistance variation ($\delta R_M$) in the device leads to a negative photoresponse for the metallic samples due to positive temperature coefficient of resistance (TCR). This resistance change can be measured with a sensitivity of one part per million via a lock-in amplifier attached to the signal out of a current pre-amplifier. Such sensitivity in measuring $\delta R_M$ implies a very large thermal sensitivity. Measurement results are then used to extract the thermal conductivity via thermal simulations.

Figure 1c-g shows a typical set of measurements from a 32 nm thick 2H-TaS$_2$ flake transferred on to the pre-patterned gold contacts with a hole of $d = 2$ μm and depth of $1$ μm etched in between using focused ion beam (FIB). Figure 1c shows the SPCM reflection map of the device taken with 200 nm/pixel step size. Corresponding photocurrent ($I_{PC}$) maps taken under 0 mV and 50 mV biases on the sample ($V_B$) in Figure 1d and e, respectively, show the local photoresponse. When no bias is applied, photoresponse results due to the electromotive force generated by the Seebeck effect at the metal-TaS$_2$ junctions. When the bias is applied, we observe photoresponse from all over the nanoflake due to the local resistance change upon laser beam heating[29]. As proposed in the previous paragraph, there is an enhancement of the absolute value of the photoresponse when the laser scans the region above the hole. Line trace taken along the crystal, through the center of the hole shows the change of local photoresponse with the laser position (Figure 1f). Corresponding measured resistance change, $\delta R_M$, can be calculated from the photocurrent, applied bias and the dark resistance ($R$) of the device: $\delta R_M \approx -R^2 \frac{I_{PC}}{V_B}$. Figure 1g shows $\delta R_M$ at each laser position. Using the $\delta R_M$ value taken at the laser position over the center of the hole, we can calculate the thermal conductivity.

Thermal distribution as a function of the position on the crystal when the laser is at the center of the hole can be calculated by solving the heat equation in two-dimensions. Under the illumination of a laser spot with a Gaussian profile, we solve the heat equation with steady-state heat flow[30]. Similar calculations for the anisotropic measurements[31] or nanowires are provided in the supporting information. We need to solve the heat equation for; $r \geq d$ and $r < d$, where $r$ is the radial distance from the center of the hole.

$$r < d \qquad \kappa \frac{1}{r}\frac{d}{dr}\left[r\frac{dT_1(r)}{dr}\right] + \frac{I\alpha}{t} e^{\frac{-r^2}{r_0^2}} = 0$$

$$r \geq d \qquad \kappa' \frac{1}{r}\frac{d}{dr}\left[r\frac{dT_2(r)}{dr}\right] - \frac{G}{t}[T_2(r) - T_0] = 0$$

Here, $\kappa$ and $\kappa'$ are the thermal conductivities of the material and the material supported by the substrate, respectively. $I$, $\alpha$, $t$, $r_0$, $T_1(r)$, $T_2(r)$, $T_0$ and $G$ are the laser power per unit area, absorbance of the crystal, thickness of the crystal, laser spot diameter, temperature distribution function for $r < d$ and $r > d$, ambient temperature and thermal boundary conductance between the crystal and the substrate from the unsuspended part of the crystal, respectively. We used volumetric Gaussian beam heating as the heat source in the equations[30]. We ignore the Newtonian cooling term as the heat loss to the air will be relatively small[32]. The general solutions for the above equations yield:

(1) $$T_1(r) = c_1 + c_2 \ln\left(\frac{r}{r_0}\right) + \frac{\alpha I r_0^2}{4\kappa t} Ei\left(\frac{-r^2}{r_0^2}\right)$$

(2) $$T_2(\gamma) = c_3 I_0(\gamma) + c_4 K_0(\gamma) + T_0$$

Here, $Ei(x)$ denotes the exponential integral, $I_0$ and $K_0$ are zero order modified Bessel functions of the first and second kind, respectively with $\gamma = r\sqrt{\frac{G}{\kappa' t}}$. To solve for the $c_n$ constants, we apply appropriate boundary conditions:

(3) $$T_2(\gamma \to \infty) = T_0$$
(4) $$\left.\frac{dT_1(r)}{dr}\right|_{r \to 0} = 0$$
(5) $$T_1(d) = T_2(\gamma)|_{r=d}$$
(6) $$\left.\kappa\frac{dT_1(r)}{dr}\right|_{r=d} = \left.\kappa'\frac{dT_2(\gamma)}{dr}\right|_{r=d}$$

The first two boundary conditions imply that the temperature far away from the center equilibrates with the ambient and the temperature under the laser spot has a finite value. The last two boundary conditions impose the continuity of the heat flow at the boundary of the suspended part of the crystal. Full solutions for $T_1(r)$ and $T_2(\gamma)$ are given in the supporting information. We assume that the lateral size of the crystal is large enough that at distances larger than $d$, sample temperature equilibrates with the substrate. Even for a material with an unusually low thermal conductivity such as 1 W/m.K, solutions to the heat equation above show that the temperature equilibrates with the substrate at the boundary of the suspended part of the crystal (see supporting information for details). For the solutions, we assume $\kappa'$ value to be similar to $\kappa$, and we used $G$ from the literature[30,32] for similar materials. However, these assumptions have no or minimal effect on the thermal distribution over the suspended part of the crystal for the aforementioned reasons and as reported for the Raman based thermal conductivity technique[10].

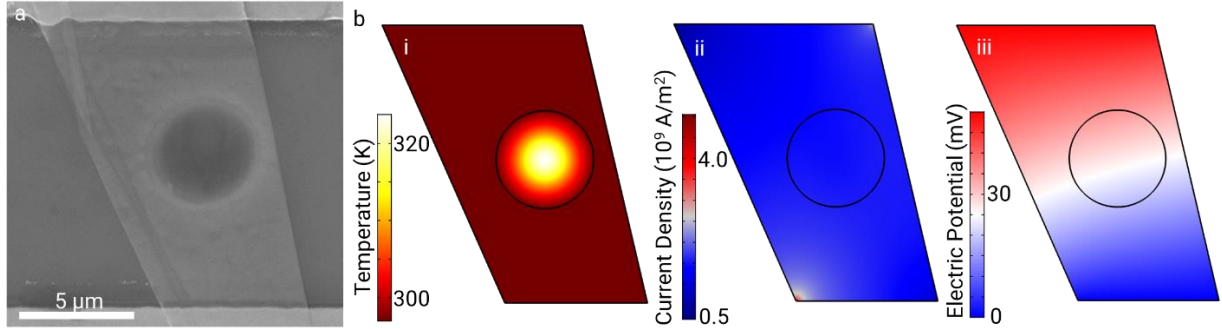

**Figure 2 a.** SEM micrograph of a 40 nm thick 2H-TaS$_2$ flake suspended over a hole of $d = 2$ μm. **b. i-**Thermal distribution over the region of the crystal between the contacts when the Gaussian laser spot is at the center of the hole. Black circle indicates the edge of the hole. **ii-** Current density distribution and **iii-** electric potential distribution between the contacts are calculated based on the local resistivity, $\rho(x,y)$.

$T(x,y)$ is used to calculate the expected resistance change, $\delta R_E$, due to photothermal heating. We can write $\rho(x,y) = \rho_0 \exp(E_A/k_B T(x,y))$ for a semiconducting sample and $\rho(x,y) = \rho_0 + \varrho[T(x,y) - T_0]$ for a metallic sample to solve for the resistance of the laser heated sample, $R_H$, numerically. It is possible to write the electrical field within the material as $\boldsymbol{E}(\boldsymbol{r}) = \rho(\boldsymbol{r})\boldsymbol{J}(\boldsymbol{r})$, and along with the continuity equation for the current density $\boldsymbol{J}$ and the Poisson's equation for $\boldsymbol{E}$ it is possible to obtain the electrical resistance[29]. However, as the crystal geometries here are typically complicated and the local current density also depends on the temperature, we used a commercially available finite element method (FEM) package (COMSOL Multiphysics) to solve for $R_H$. For simpler geometries and materials with low electrical resistance the analytical solutions can be used. Figure 2a shows an SEM image of a typical device and Figure 2b shows the corresponding FEM simulations of the thermal distribution, the current density distribution and the electric potential over the sample. The expected resistance change can be obtained by subtracting the measured dark resistance from the calculated $R_H$, $\delta R_E = R_H - R$. We can match the values of $\delta R_E$ with the measured resistance change, $\delta R_M$, by using $\kappa$ as the fitting parameter in the temperature distribution function.

To demonstrate the applicability of our method, we measured the thermal conductivity of V$_2$O$_3$ nanoplates. V$_2$O$_3$ is an exemplary correlated oxide with a known thermal conductivity in its metallic state and shows a little variation in its properties from bulk to thin sheets[33]. Thus, we used V$_2$O$_3$ as a test sample to check the validity of the proposed thermal conductivity measurement method The synthesis of the nanoplates are discussed elsewhere[33]. These V$_2$O$_3$ nanoplates are synthesized over sapphire substrates with thicknesses ranging from a few to a few hundred nanometers and can be transferred on other substrates using polymer assisted transfer techniques. Once the crystals are transferred on to a FIB drilled sapphire substrate, they are in the paramagnetic metallic (PM) state and the thermal conductivity of the PM phase has been reported as 4.5 W/m.K at room temperature[34]. We preferred a V$_2$O$_3$ crystal with the thickness of ~130 nm to make a better comparison with the literature. Absorption coefficient for V$_2$O$_3$ is determined from an earlier report[35]. We measured the thermal conductivity as $4.5 \pm 1.0$ W/m.K

(see supporting information for the details). In our measurements, we ignored the heating by the reflected light from the bottom of the hole. Our measurements are in an excellent agreement with the value reported in the literature. Thermal simulations show that the crystal under the laser spot heats up by ~0.2 K/µW during the measurements. This implies that our method can be used to measure the thermal conductivity of nanosheets in the vicinity of the thermally induced phase transitions observed in materials such as $V_2O_3$.

To further illustrate the applicability of the bolometric thermal conductivity measurement method, we measured the thermal conductivity of 2H-TaS$_2$ flakes. 2H-TaS$_2$ is an intriguing van der Waals layered material that displays superconductivity[36] at 0.5 K and charge density wave (CDW) transition[37,38] around 75 K. The superconducting transition temperature increases from 0.5 to 2.2 K as the number of layers decrease[39]. Thermal conductivity of 2H-TaS$_2$ has not been measured to date. We fabricated four devices with similar crystal thicknesses and measured the thermal conductivity of each crystal. Once the thermal conductivity measurements are finished, we measured the crystal dimensions using AFM. The thermal profile is calculated from the temperature dependent resistivity of 2H-TaS$_2$, $\rho(T) = 0.4(0.2\,)m\Omega.cm(1 + 0.0025K^{-1}[T - T_0])$ [40,41]. This relation holds down to the onset of the CDW transition. Average in-plane thermal conductivity of 2H-TaS$_2$ is measured as $13.2 \pm 1.0$ W/m.K (Table S1 in the supporting information shows the detailed parameters for all the measurements). The absorption coefficient ($\alpha$) of 2H-TaS$_2$ measured for each device we fabricated before transferring the crystals on to the gold contacts. Figure 3a shows the change of absorption coefficient with the crystal thickness for 2H-TaS$_2$ at 642 nm. Radius of the Gaussian beam ($r_0$) is extracted from first derivative of the intensity with respect to the laser position at the edge of the gold contacts (Figure 3b). Figure 3c shows the calculated change in the device resistance for various $\kappa$ and $G$ values. Even for very low thermal conductivity materials, a large range of thermal boundary conductance values give accurate $\delta R_E$. Thermal conductivity measurements taken with 532 nm laser (see supporting information) yields the same thermal conductivity within the error margin.

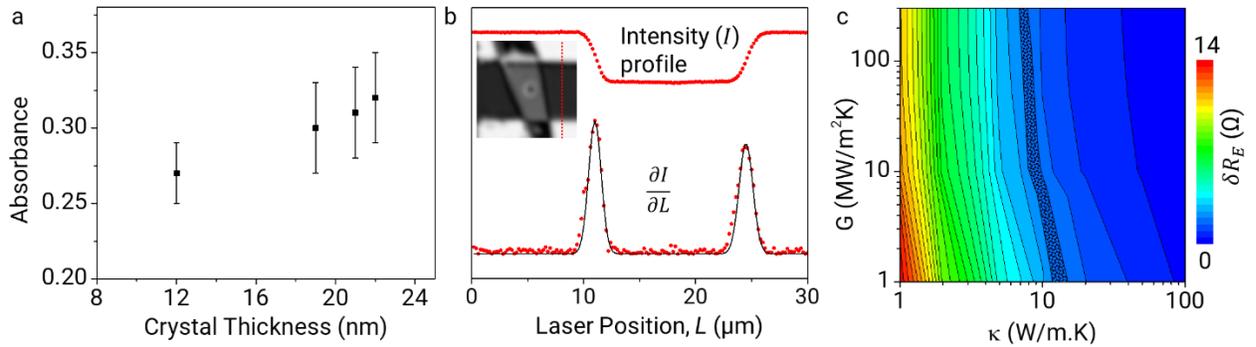

**Figure 3 a.** Absorbance vs. crystal thickness measured for four different 2H-TaS$_2$ crystals of various thicknesses. **b.** Reflected light intensity ($I$) profile extracted from the reflection map shown in the inset is used to calculate the Gaussian beam radius. $\frac{\partial I}{\partial L}$ is the derivative of light intensity with respect to the laser position, $L$. The Gaussian fit to the profile gives the radius, $r_0$, of the focused laser spot. **c.** Contour Plot shows how $\delta R_E$ changes for various set of $G$ and $\kappa$ values. Shaded area indicates the $\kappa$ values for $\delta R_E = \delta R_M$. It is evident that the variation in $\kappa$ due to $G$ is ~10% for a very large range of $G$ values.

To support the bolometric thermal conductivity measurements of 2H-TaS$_2$, its electronic and thermal properties are obtained via the first-principles calculations based on density functional theory (DFT)[42,43] as implemented in the Vienna *ab initio* simulation package (VASP)[44,45] (see methods for further details). Ground state geometry (Figure 4a) of 2H-TaS$_2$ is obtained and the calculated lattice constants (a$_{calc.}$=3.31 A and c$_{calc.}$=12.07 A) are in well agreement with the experimental data (a$_{exp.}$=3.32 A and c$_{exp.}$=12.10 A)[46]. Following the structural optimization, the electronic band structure and the phonon spectrum is calculated as shown in Figure 4b and c, respectively. In line with experimental results, 2H-TaS$_2$ has a metallic character and degeneracy at high symmetry points is altered with inclusion of spin-orbit coupling. All phonon modes are real indicating the structural stability[47].

Relaxation time and the density of electrons must be specified to determine the in-plane thermal conductivity. Relation between the mobility and the relaxation time is given by the following equation, $\mu = \frac{\tau e}{m_e}$, where $e$ is the electron charge, $m_e$ is the mass of electron and $\tau$ is the relaxation time. By considering the room temperature in-plane mobility data reported in literature[48], $\tau$ is specified as 5.69 x 10$^{-15}$ s which is reasonable based on the Drude theory of metals. The density of electrons ($n$) can be determined from the experimental measurements of Hall coefficient ($R_H$)[49], $R_H = \frac{1}{ne}$ and calculated as 3.13 x 10$^{22}$ cm$^3$ at room temperature. Accordingly, electronic thermal conductivity ($\kappa_e$) using the calculated electronic density and relaxation time is estimated as 4.73 W/m.K. The lattice thermal conductivity ($\kappa_l$) is determined as 8.62 W/m.K and 6.81 W/m.K by using the iterative solution and relaxation time approximation methods, respectively. Therefore, the total thermal conductivity of 2H-TaS$_2$ at room temperature is determined to be in the range of 11.55-13.36 W/m.K. This range is in an excellent agreement with our experimental measurements.

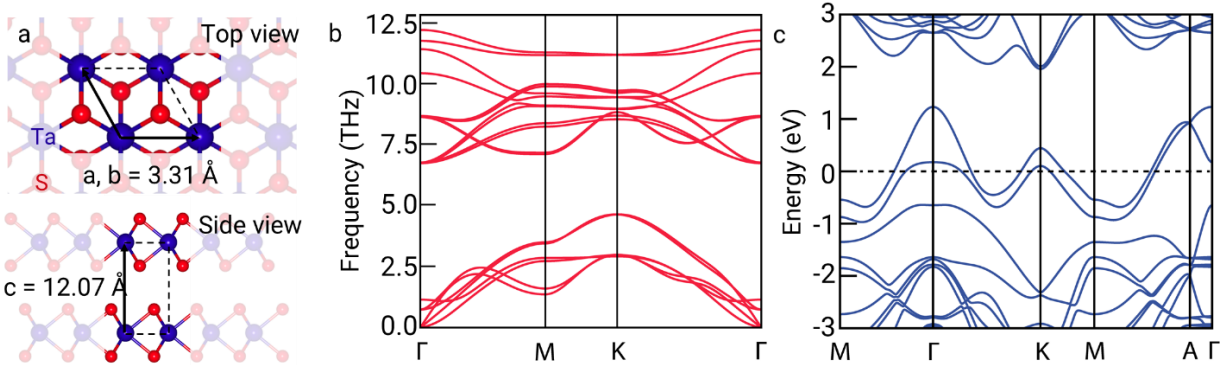

**Figure 4 a.** Top and side views of 2H-TaS$_2$ crystal structure is depicted with 2H layers are stacked together and the lower layer is rotated by 60° with respect to the upper layer. Ta and S atoms are marked on the figure. **b.** All positive phonon modes reveal the dynamical stability of 2H-TaS$_2$ and provide a basis (harmonic force constants) to determine lattice thermal conductivity together with the anharmonic force constants. **c.** Electronic band structure with inclusion of spin-orbit coupling along with high symmetry points indicate the metallic character of the system and band dispersions form a basis for the electronic thermal conductivity.

Now, we would like to discuss the applicability of our method on semiconducting nanosheets. Our method relies on the precise measurement of the electrical resistance variation upon the laser heating. The change in the electrical resistance over the suspended part of the crystal due to the light induced heating must be differentiated from the other photoresponse mechanisms prevalent in semiconductors. As the photoresponse in semiconductors may have multiple reasons, applicability of the bolometric thermal conductivity measurement technique on semiconducting nanosheets and nanowires requires a deeper analysis of the measurements. For the timescales shorter than a millisecond, photoconductivity in a semiconductor under bias can result from the formation of non-equilibrium carriers due to light absorption, separation of non-equilibrium carriers due to built-in electric fields or photothermal effects[50]. Furthermore, the strain induced bandgap changes within the suspended region will create built-in electric fields and will further complicate the analysis of the data. We attempted measuring the thermal conductivity of 2H-$MoS_2$ and 2H-$WS_2$ few layer crystals mechanically exfoliated from the bulk using a sticky tape. Both materials are exemplary layered TMDCs with direct bandgaps in the monolayer and become indirect gap semiconductors in the bulk. The exfoliated crystals are transferred over the holes etched on sapphire and indium needles are drawn on to the crystals at elevated temperatures as top contacts to minimize the contact resistance. For a 12 nm thick $WS_2$ sample we measured the thermal conductivity of flake as 8 W/m.K by assuming that the enhancement of the photocurrent over the hole is entirely due to the bolometric effect (see supporting information for details). This value is smaller than what has been reported previously (12 W/m.K)[51] possibly due to the aforementioned reasons. In some samples we observed formation of multipolar junction like photoresponse that can be attributed to the strain induced changes in the charge doping over the suspended part of the crystal[52,53]. This manifests itself as multiple peaks in the photoresponse under bias and complicates the extraction of the bolometric effect (see supporting information). Further investigation is needed to elucidate the usability of the bolometric thermal conductivity measurement on semiconducting nanosheets.

Our method is very similar to the Raman thermometry in terms of the measurement errors and limitations[54,55]. Local temperature measurements both in Raman thermometry and our method relies on modeling of the temperature distribution over the suspended part of the crystal with $\kappa$ being a fitting parameter. Our method is applicable at any temperature if the resistivity of the material varies with the temperature. Phenomena due to the electronic correlations that results in abrupt changes in the electrical resistivity would jeopardize our measurement method in the close vicinity of the phase transitions, yet this limitation applies to all thermal conductivity measurement techniques. Moreover, since the residual resistance for the metals at very low temperatures have very weak temperature dependence, our method would fail at such regimes as well. Another problem associated with the bolometric measurement method we introduce would be the large contact resistance[29]. When the contact resistance dominates the total resistance of the device, bolometric response is significantly reduced.

One of the major advantages of our method is the high sensitivity of the measurements. Especially for the materials with large $|\varrho|$ values, laser power as small as ~µW produces a measurable photoresponse. 0.2 K average temperature rise under the laser spot can increase the electrical resistance by a few mΩ and this change is easily measurable in a ~100 Ω crystal. As a comparison with the Raman thermometry-based method, typical first-order linear temperature coefficients of the Raman modes are in the range of ~0.005 to 0.02 $cm^{-1}$/K. Even for a long-focal-length spectrometer equipped with a cutting-edge charge coupled device, the resolution is ~0.5 $cm^{-1}$ for the visible light. Thus, the minimum average temperature rise of 25 - 100 K over the sample is required for a reliable measurement. This is particularly important for the temperature dependent study of the thermal conductivity especially in the vicinity of thermally induced phase transitions. Moreover, oxidation or sample degradation due to laser heating is minimized in our method. Another advantage of our method is the relative simplicity of the measurement setup. Although we used an SPCM for the measurements, a laser coupled to an optical microscope could be used to perform similar measurements. Finally, the method is also applicable to nanowires and materials with anisotropic in-plane thermal conductivity with a suitable choice of the laser shape.

In summary, we introduced a novel bolometric effect based thermal conductivity measurement method that can be applied to nanosheets and nanowires with temperature dependent electrical resistivity. As a demonstration of the method, we measured the room temperature thermal conductivity for $V_2O_3$ nanosheets and showed that the measured value is comparable to the previous reports. We measured the room temperature thermal conductivity of 2H-$TaS_2$ as 13.2±1.0 W/m.K for the first time and performed *ab initio* calculations to find its thermal conductivity numerically. We discussed the versatility of our technique in detail and showed that it is superior to other commonly used methods in terms of the thermal sensitivity. Accuracy and applicability of our method is comparable to Raman thermometry, yet, with much higher thermal sensitivity. As a final remark, our technique can be extended to the scanning thermal microscopy. Although we used a laser beam as the heat source, same measurement can be performed using a heated scanning probe instead. This could eliminate the need for the measurement of $\alpha$ and as with the precise positioning of the scanning probe is possible, a better modelling of the thermal distribution could be performed to increase the accuracy.

**Methods**

SPCM measurements are performed using a commercial setup (LST Scientific Instruments) under 642 nm illumination unless otherwise stated.

For the *ab initio* calculations, the exchange-correlation interactions were estimated by generalized gradient approximation (GGA) with inclusion of spin-orbit coupling[56,57]. The van der Waals (vdW) interactions were taken into account by using Grimme method[58,59]. The element potentials described by projector augmented wave (PAW)[60,61] method with a kinetic energy cutoff of 450 eV. The Brillouin zone was sampled with 17x17x3 k-point mesh by using Monkhorst-Pack grids[62]. The energy convergence for ionic and electronic relaxations was set to $10^{-6}$ eV whereas the maximum force allowed on atoms is less than $10^{-4}$ eV/A$^{-1}$.

The electronic thermal conductivity was calculated by solving semi-classical Bolztmann transport equation (BTE) considering constant relaxation time and the rigid band approximation[63]. The lattice thermal conductivity was determined by iteratively solving BTE equation where zeroth iteration solution corresponding to the Relaxation Time Approximation (RTA)[64,65]. Harmonic and anharmonic force constants were calculated by using finite displacement method[66].

**Author Contributions**

T.S.K. proposed the method and conceded the experiments. O.Ç. and N.M. contributed equally to the work. O.Ç. developed device fabrication recipes, performed the measurement analysis and simulations, N.M. performed the SPCM measurements and device fabrications with the help from

# Supporting Information: Thermal Conductivity Measurements in Nanosheets and Nanowires Using Photothermal Electrical Resistance Change

Onur Çakıroğlu[2], Naveed Mehmood[1], Mert Miraç Çiçek[1], Aizimaiti Aikebaier[1], Hamid Reza Rasouli[1], Engin Durgun[1], T. Serkan Kasırga[1,2]

[1] Bilkent University UNAM – National Nanotechnology Research Center, Ankara, Turkey 06800

[2] Department of Physics, Bilkent University, Ankara, Turkey 06800

**Thermal Conductivity Measurements on 2H-TaS$_2$ Flakes**

To confirm the applicability of our method at a different laser wavelength, we measured the thermal conductivity of 2H-TaS$_2$ using a 532 nm DPSS laser. Figure S1 shows the scanning photocurrent measurements taken on a 19 nm thick crystal transferred over metallic contacts.

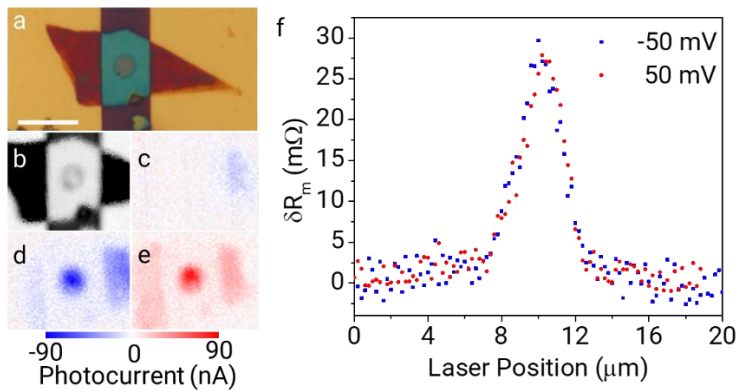

**Figure S1 a.** Optical microscope micrograph of 2H-TaS$_2$ crystal transferred over gold contacts with a hole etched in between. The scale bar is 10 µm. **b.** Reflection map, **c.** 0 mV, **d.** 50 mV and **e.** -50 mV photocurrent maps. **f.** $\delta R_m$ vs. the laser position plotted along the width of the measurement for 50 mV and -50 mV biases.

For the $\delta R_m$ values measured with 10 µW laser power, we performed COMSOL simulations and find the thermal conductivity as $\kappa = 14.0 \pm 1.0$ W/m.K . As mentioned in the main text one of the major advantages of our technique over the other thermal conductivity measurement methods is the sensitivity of detection. With 10 µW laser power, the maximum temperature in the middle of the hole is 2.7 K above the ambient temperature. Considering the signal to noise ratio in our measurements even with a 2 µW laser power it is possible to extract the thermal conductivity, that corresponds to a mere ~0.5 K maximum temperature rise at the center of the suspended part.

**Nanowires & Anisotropic Measurements**

The method we proposed in the main text can be used on nanowires and materials with anisotropic in-plane thermal conductivity. The measurement on nanowires requires no change to the experimental setup given in the main text. The laser beam parked at the suspended part of the nanowire will create a thermal gradient from the laser point. Then one-dimensional heat transport equation is solved for the thermal profile. With the known dimensions of the nanowire it is straightforward to calculate the resistance of the nanowire with and without the laser illumination.

Anisotropic thermal conductivity measurements require an aperture on the laser beam to create a focused line rather than a spot. The anisotropic crystal suspended over a trench can be oriented along certain crystal axes and with the line shaped laser beam on the sample, the problem is effectively reduced to solving one dimensional heat transport equation.

Here, again we ignore the Newtonian cooling term and have two equations for the suspended and the supported parts of the crystals:

$$\kappa \frac{d^2}{d^2 x}[T_1(x)] + \frac{I\alpha}{t} e^{\frac{-x^2}{w_0^2}} = 0 \text{ for suspended section}$$

$$\kappa' \frac{d^2}{d^2 x}[T_2(x)] - \frac{G}{t}(T_2(x) - T_0) = 0 \text{ for the supported section}$$

where $w_0$ is the full width half maximum of the laser line and the other parameters represent the same quantities as in the main text. The general solution to the above equations yield:

$$T_1(x) = C_1 + C_2 x - \frac{\alpha \sqrt{\pi} w_0}{2\kappa} \left( \exp\left(-\frac{x^2}{w_0^2}\right) \frac{w_0}{\sqrt{\pi}} + \text{erf}\left(\frac{x}{w_0}\right) x \right)$$

$$T_2(x) = T_0 + C_3 \exp\left(x \sqrt{\frac{G}{t\kappa'}}\right) + C_4 \exp\left(-x \sqrt{\frac{G}{t\kappa'}}\right)$$

where erf is the error function. When we use the appropriate boundary conditions, we find that $C_2 = C_3 = 0$ and

$$C_1 = T_0 + C_4 \exp\left(-w \sqrt{\frac{G}{t\kappa'}}\right) + \frac{\alpha \sqrt{\pi} w_0}{2\kappa} \left( \exp\left(-\frac{x^2}{w_0^2}\right) \frac{w_0}{\sqrt{\pi}} + \text{erf}\left(\frac{w}{w_0}\right) w \right)$$

$$C_4 = \frac{\alpha w_0}{2} \sqrt{\frac{t\pi}{G\kappa'}} \frac{\text{erf}\left(\frac{w}{w_0}\right)}{\exp\left(-w \sqrt{\frac{G}{t\kappa'}}\right)}$$

Here $w$ is the width of the trench.

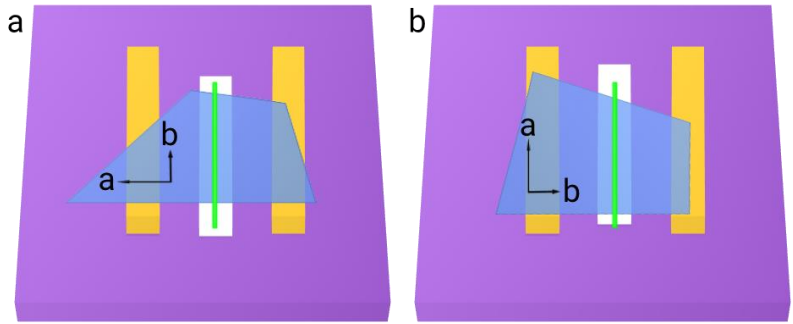

**Figure S2 a.** Illustration of the anisotropic measurement configuration. SiO$_2$ substrate with a trench at the center has two gold colored electrical contacts and a crystal is suspended over the trench with contacting the crystals. Green line going through the middle depicts the laser line illuminating the sample. Two different crystal axes are marked on the blue nanosheet. **b.** Same configuration as in **a** with different crystal orientation.

**Solutions to $T_1(r)$ and $T_2(\gamma)$ for Isotropic Nanosheets**

Under appropriate boundary conditions stated in the main text, the coefficients in eq. (1) and (2) of the main text becomes:

$$c_1 = T_0 + \frac{\alpha P K_0(\gamma_R)}{2\pi d t K_1(\gamma_R)}\sqrt{\frac{t}{G\kappa'}}\left[1 - e^{-\frac{d^2}{r_0^2}}\right] + \frac{\alpha P}{2\pi\kappa t}\left[2\ln(d) - Ei\left(-\frac{d^2}{r_0^2}\right)\right]$$

$$c_2 = -\frac{\alpha r_0^2}{2\kappa}$$

$$c_3 = 0$$

$$c_4 = \frac{\alpha r_0^2}{2 d K_1(\gamma_R)}\sqrt{\frac{t}{G\kappa'}}\left[1 - \frac{e^{-d^2}}{r_0^2}\right]$$

Where, $K_0$ and $K_1$ are the second kind Bessel functions of the zeroth and the first order, $Ei$ is the exponential integral. Here, $P$ is the laser power ($I = \frac{P}{\pi r_0^2}$), $\gamma_d = d\sqrt{\frac{G}{\kappa' t}}$ and the other symbols are given in the main text.

**Temperature at the Boundary of the Hole with respect to the Thermal Conductivity**

As explained in the main text, both $G$ and $\kappa'$ doesn't affect the measured thermal conductivity as even for very low thermal conductivity values, the temperature rapidly equilibrates with the environment. Figure S3 shows the temperature at the boundary of a sample for various thermal conductivity values at various $G$ values. We assumed $\kappa' = 0.9\kappa$.

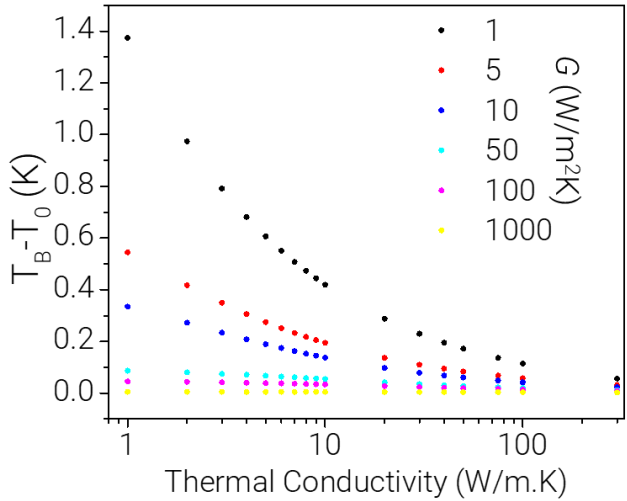

**Figure S3** Figure shows change in the boundary temperature with respect to the ambient temperature at various thermal conductivity and thermal boundary conductance values. Data is generated for a sample under 10 µW laser illumination with 0.33 absorption. Even for materials with very low thermal conductivity and very low thermal boundary conductance, the temperature at the boundary only increases a few Kelvin above the ambient temperature.

**Thermal Conductivity Measurement on $V_2O_3$**

Thermal conductivity measurement on a 130 nm thick $V_2O_3$ crystal is performed on a sample transferred over a hole etched on a sapphire substrate. Indium contacts are later drawn to make the electrical connection with the crystal. Here, we used indium contacts are used to reduce the contact resistance as we used thick crystals. SPCM taken at 9.5 µW laser power results in a bolometric response over the hole. Thermal simulations reveal the thermal conductivity of the material as 4.5 W/m.K for biases ranging from -200 mV to 200 mV with 50 mV steps. Absorption coefficient is used as 0.8.

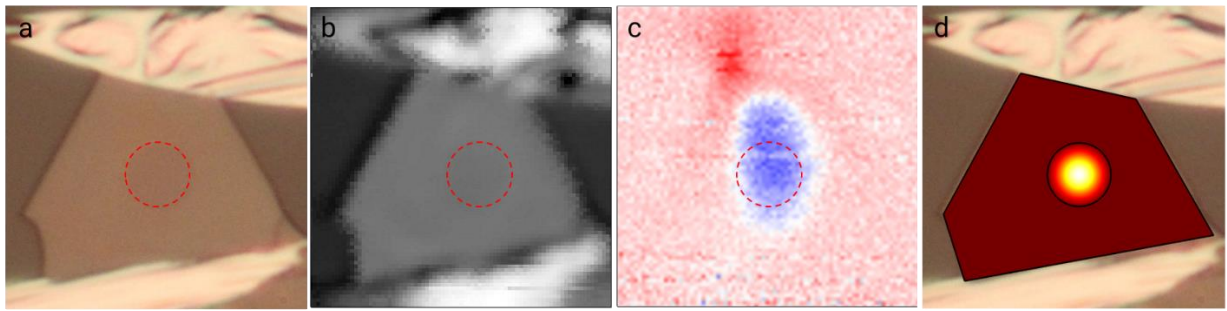

**Figure S4 a.** Optical micrograph of $V_2O_3$ crystal. The dashed red circle denotes the position of the hole. **b.** Reflection map and **c.** the photocurrent map. Slight elongation of the bolometric response towards the contact is possibly due to an air pocket trapped under the crystal. **d**. FEM result of the thermal map is overlaid on the optical microscope image. The hottest point under 9.5 µW illumination is ~2.5 K above the ambient temperature.

**Absorbance Measurements**

To measure the absorbance of the crystals, $2H\text{-}TaS_2$ crystals are first exfoliated over a PDMS stamp. Then, the absorbance is measured via a spectrometer. First, the laser intensity, $I_0$, is measured with nothing in the path. Then, the transmitted intensity, $T_0$, and the reflected intensity, $R_0$, of the bare PDMS substrate are measured in the vicinity of the crystal of interest. Finally, the transmitted, $T$, and reflected, $R$, intensities over the crystal are measured. Using these five measured quantities we calculated the absorbed intensity for each crystal, $\alpha = \frac{I_0 - (T+R) - (T_0+R_0)}{I_0}$. Once the crystals are transferred on the devices, atomic force microscopy is used to measure the thicknesses to correlate with the absorbed laser intensity.

**List of Measured Thermal Conductivities of $2H\text{-}TaS_2$ Samples**

In the main text, we reported the average value measured for all $TaS_2$ crystals with different thicknesses. Table S1 is a list of samples we measured with relevant material parameters. The thermal conductivity measurements from each sample is the average value obtained under 50 mV and -50 mV biases.

There is no trend in thickness dependence in our measurements. However, considering that the thicknesses of our crystals are not too different from each other, we don't expect a difference. Sample to sample variation in the thermal conductivity values is within the error and consistent with the DFT calculations.

**Table S 1** Table shows the values measured for each 2H-TaS$_2$ crystal. Thermal conductivity values for each individual sample are an average of measurements various biases.

| Sample # | Measurement Wavelength (nm) | Thickness (nm) | Absorption Coefficient (α) | Laser Power (μW) | Dark Resistance (Ω) | $\rho(T)$ (mΩ.cm) | $\delta R$ (Ω) | κ(W/m.K) |
|---|---|---|---|---|---|---|---|---|
| 1 | 642 | 34.0 | 0.33 | 40 | 124 | 0.60 $(1 + 0.0025 K^{-1}[T - T_0])$ | 0.019±0.001 | 12.6 ± 0.7 |
| 2 | 532 | 19.1 | 0.28 | 10 | 131 | 0.34 $(1 + 0.0025 K^{-1}[T - T_0])$ | 0.027±0.001 | 14.0 ± 1.0 |
| 3 | 642 | 32.1 | 0.33 | 211 | 138 | 0.25 $(1 + 0.0025 K^{-1}[T - T_0])$ | 0.48±0.01 | 14.0 ± 1.0 |
| 4 | 642 | 33.1 | 0.33 | 211 | 267 | 0.52 $(1 + 0.0025 K^{-1}[T - T_0])$ | 0.75±0.01 | 12.0 ± 1.0 |
| | | | | | | | **Average** | 13.2 ± 1.0 |

**Bolometric Thermal Conductivity Measurements on WS₂ and MoS₂ Thin Flakes**

As discussed in the main text, in principle it is possible to measure semiconducting nanosheets and nanowires using our measurement method. The main challenge in semiconducting crystals is to distinguish the various photoconductance mechanisms from the photothermal response. For one WS₂ and one MoS₂ device, we were performed scanning photocurrent microscopy to extract the thermal conductivity. Figure S5 shows a 12 nm thick WS₂ flake exfoliated from the bulk with a sticky tape and transferred on to a hole drilled on sapphire substrate. Then, indium contacts are drawn on to the crystal for electrical contacts. The SPCM under 400 mV reveals a positive photoresponse at the center of the hole. By using the measured resistance change at the center of the crystal and the Arrhenius relation measured for the crystal ($\rho(T) = \rho_0 e^{\frac{80\ \text{meV}}{k_B T}}$) we calculated the thermal conductivity of the sample. Here we used the exact relation to calculate the resistance change as it is comparable to the resistance of the sample: $-\delta R_M = \frac{R^2 I_{PC}}{V_B + R I_{PC}}$.

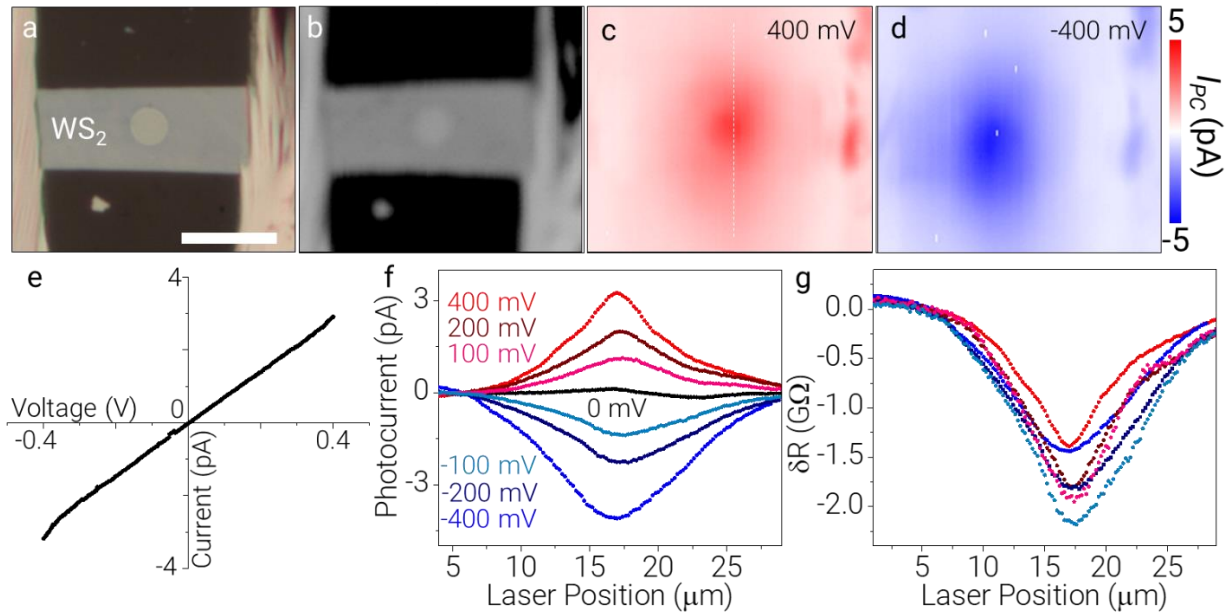

**Figure S5 a.** Optical microscope micrograph of an indium contacted WS₂ crystal. Scale bar is 10 μm. **b.** Corresponding intensity map of the reflection image. White dashed line represents the line cuts taken at different biases. **c.** and **d.** show photocurrent maps at 400 and -400 mV biases. **e.** IV curve of the device shows an Ohmic response within the measured bias interval. **f.** Photocurrent line traces taken at different biases parallel to the contacts as shown in **b**. **g.** Measured resistance change vs. the laser position show for different biases. Same coloring is used as in **f**. There is a slight increase in the resistance change towards the lower biases. This may hint existence of a different mechanism at play for the photoconductance.

MoS₂ devices yielded a much more interesting photoresponse. A crystal of 10 nm in thickness is placed on a hole etched in sapphire and contacted with indium. SPCM reveals a symmetric bipolar response over the hole. When a bias is applied, there are two peaks appearing around the hole. The zero bias response hints for the existence of another mechanism and the effect of the bolometric effect requires a further analysis of the data.

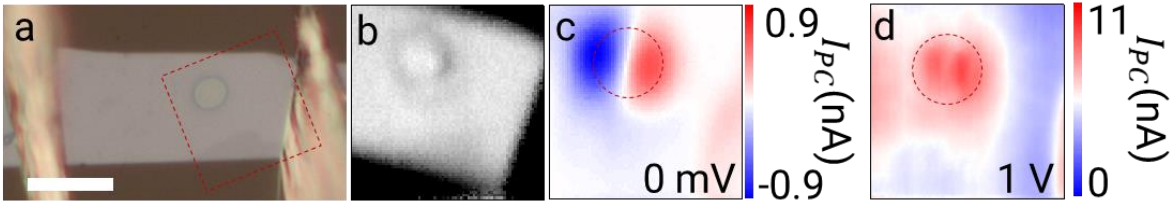

**Figure S 6 a.** Optical microscope micrograph of the MoS$_2$ device. The dashed square shows the region where the SPCM is taken from. Scale bar is 10 μm. **b.** Reflection map and **c.** photocurrent map under zero bias. **d.** Photocurrent map under 1 V bias. Dashed circle shows the position of the underlying hole.

**Sample Characterization**

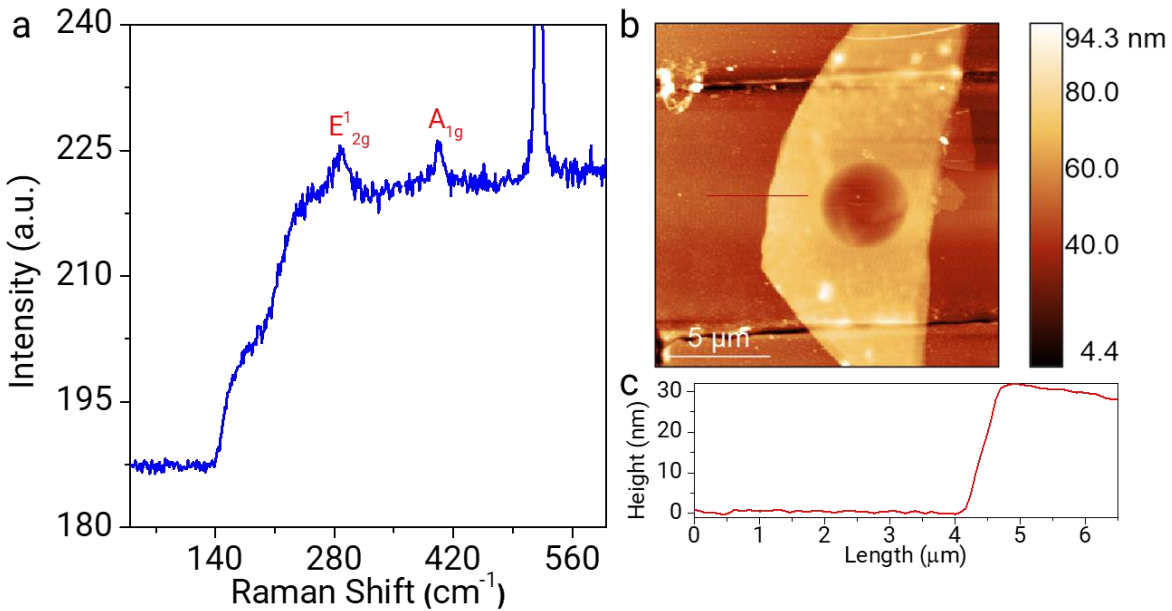

**Figure S7 a.** Raman spectrum from a TaS$_2$ crystal. **b.** AFM height map of device 3 and **c.** height trace taken along the red line. Same set of data is taken for all devices.